\documentclass[12pt,preprint]{aastex}

\usepackage{lscape}

\shorttitle{The Carina Tangent Field}
\shortauthors{\shortauthors{Kaltcheva \& Golev}}

\begin{document}

\def\Stromgren{Str{\"o}mgren }
\def\o{\"o}
\def\Lod{Lod$\acute{\mathrm e}$n}
\def\V0{$V_0$}
\def\V{$V$}
\def\Eby{$E(b-y)$ }
\def\EBV{$E(B-V)$ }
\def\Mv{$M_V$ }
\def\Av{$A_V$ }
\def\b-y{$b-y$ }
\def\c1{$c_1$ }
\def\m1{$m_1$ } 
\def\BrM1{$[m_{1}]$ }
\def\BrC1{$[c_{1}]$ }
\def\d{$^\circ$}
\def\'{$ ^{\rm '}$ }
\def\C0{$c_0$ }
\def\M0{$m_0$ }
\def\by0{$(b-y)_0$ }
\def\Hb{$H\beta$ }
\def\uvbyb{$uvby\beta$ }
\def\BV0{$(B-V)_0$ }
\def\Vlsr{$V_\mathrm{lsr}$ }

\title{Galactic Structure Toward the Carina Tangent}


\author{N. T. Kaltcheva}
\affil{Department of Physics and Astronomy, University of Wisconsin Oshkosh,
  800 Algoma Blvd., Oshkosh, WI 54901, USA}
\email{kaltchev@uwosh.edu}

\and

\author{V. K. Golev}
\affil{Department of Astronomy, Faculty of Physics, St Kliment Ohridski University of Sofia, 
 5 James Bourchier Blvd., BG-1164 Sofia, Bulgaria}
\email{valgol@phys.uni-sofia.bg}


\begin{abstract}
This investigation presents a photometric study of the Galactic structure
toward the Carina arm tangent. The field is located between $280^\circ$ and
$286^\circ$ galactic longitude and $-4^\circ$ to $4^\circ$ galactic latitude.
All currently available \uvbyb\ data is used to obtain homogeneous color
excesses and distances for more than 260 stars of spectral types O to G.  We
present revised distances and average extinction for the open clusters and
cluster candidates NGC 3293, NGC 3114, \Lod\ 46 and \Lod\ 112. The cluster
candidate \Lod\ 112 appears to be a very compact group at a true distance
modulus of $11.06\,\pm\,0.11$(s.e.) ($1629^{+84} _{-80}$ pc), significantly
closer than previous estimates. We found other OB stars at that same
distance and, based on their proper motions, suggest a new OB association at
coordinates $282^\circ < l < 285^\circ$, $-2^\circ < b < 2^\circ$.  Utilizing
BV photometry and spectral classification of the known O-type stars in the
very young open cluster Wd~2 we provide a new distance estimate of
$14.13\,\pm\,0.16$ (s.e.) ($6698^{+512} _{-475}$ pc), in excellent agreement
with recent distance determinations to the giant molecular structures in this
direction. We also discuss a possible connection between the H{\sc ii} region
RCW 45 and the highly-reddened B+ star CPD $-$55~3036 and provide a revised
distance for the luminous blue variable HR Car.
\end{abstract}

\keywords{Galaxy: structure; open clusters and associations: individual
  (NGC 3293); open clusters and associations: individual (\Lod\ 46); open
  clusters and associations: individual (\Lod\ 112); open clusters and
  associations: individual (NGC 3114); open clusters and associations:
  individual (Wd 2); stars: distances; stars: individual (HR Car)}

\section{Introduction}

In the fourth Galactic quadrant, the direction toward $282^\circ$-$285^\circ$
galactic longitude corresponds to maxima in the thermal radio continuum, H{\sc
  i} and CO emissions \citep{tay93, blo90, bro92}. Other kinds of tracers also
indicate that in this direction the line of sight is tangential to a large
segment of the Carina arm.  Based on a multiwavelenght study of star-forming
complexes, it is  found that longitudes $283^\circ$-$284^\circ$
correspond to a tangential direction in both the 3- and 4-arm models of the
grand design of the MW  \citep{rus03, val08}.

The OB stars observed in the $283^\circ$-$284^\circ$ longitude range  delineate a
sharp outer edge of the Carina arm at about 2-3 kpc from the Sun \citep{gra70,kal10}. The field located between the associations in Vela ($262^\circ < l <
268^\circ$) and Car~OB1 ($284^\circ < l < 288^\circ$) is not known to be
dominated by any prominent OB association \citep{hum78,mel95}. The only known
SNR in the region is G279.0+1.1 \citep{dun95}.

Local and intermediate-scale features of the Galactic disk, like arm-splitting
and branching, are important part of the grand design of the Milky Way (see
\citet{rus03} for a recent discussion). This paper is focused on a field
between $280^\circ$ and $286^\circ$ in the Galactic disk with aim to address
the present deficiency in the study of the structure toward the edge of the
Carina arm.  Our study is based on \uvbyb\ photometric distances and provides
new insights on the apparent groupings and layers in this region.

\section{The sample}
\begin{figure*} 
\begin{center}
\includegraphics[scale=.60]{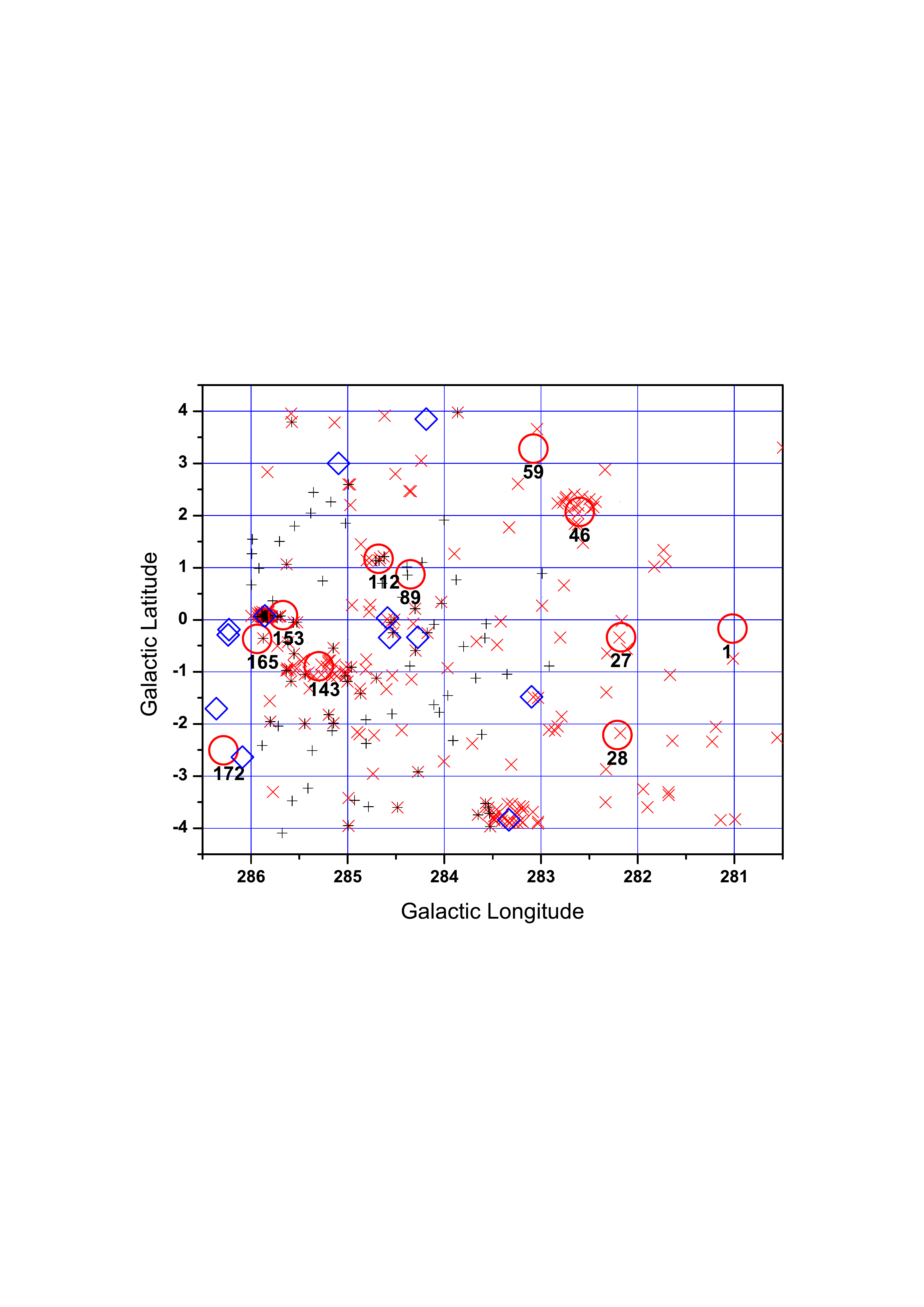}
\caption[]{All stars with \uvbyb photometry  plotted in Galactic
  coordinates (longitude and latitude, in degrees). Crosses indicate stars from \citet{hau98} and plus symbols -
  stars from \citet{kal03}. All \Lod\ cluster candidates are shown with large
  open symbols and are labeled. The known clusters in the field
  \citep{dia03} are shown with large diamonds (unlabeled; see Table
  2).\label{fig1} See the electronic edition of the PASP for a color version
  of this figure.}
\end{center}
\end{figure*}

The field under consideration is located between $280^\circ$ and
$286^\circ$ galactic longitude and $-4^\circ$ and $4^\circ$ galactic
latitude.  All \uvbyb\ data within this coordinate range was extracted from
the catalog of \citet{hau98} and combined with the \uvbyb\ photometry from
the catalog of \citet{kal03}.  The sample contains more than 300 stars of
spectral types O to G with complete \uvbyb\ photometry and is listed in
Table~1, available only electronically. Fig. \ref{fig1} presents these stars
plotted in Galactic coordinates. All known open clusters and all
\Lod\ cluster candidates are shown as well and are listed in Table~2\setcounter{table}{2}.

\section{Calculation of interstellar extinction and distances}
\begin{figure*} 
\begin{center}
\includegraphics[width=15pc]{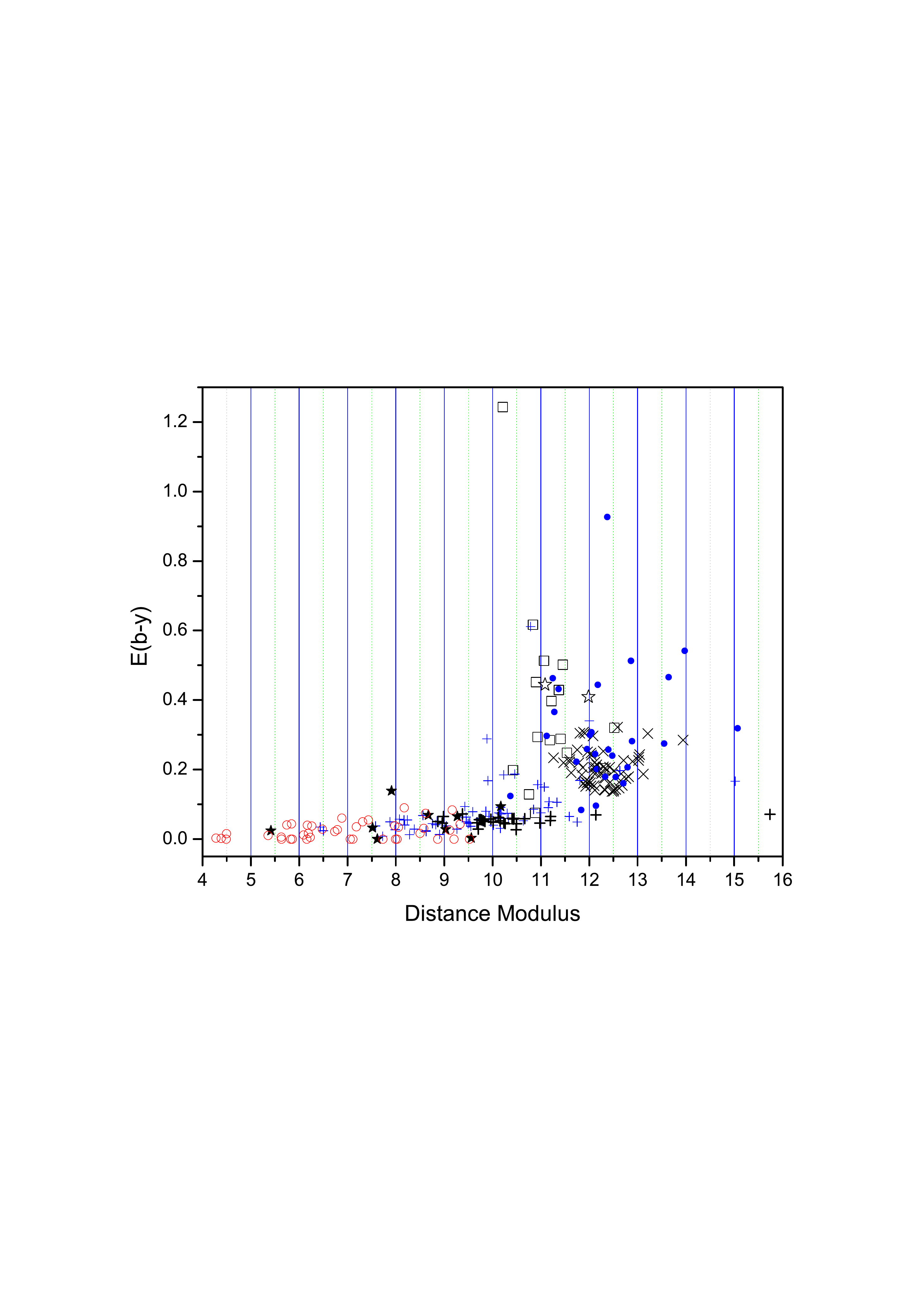}
\includegraphics[width=15pc]{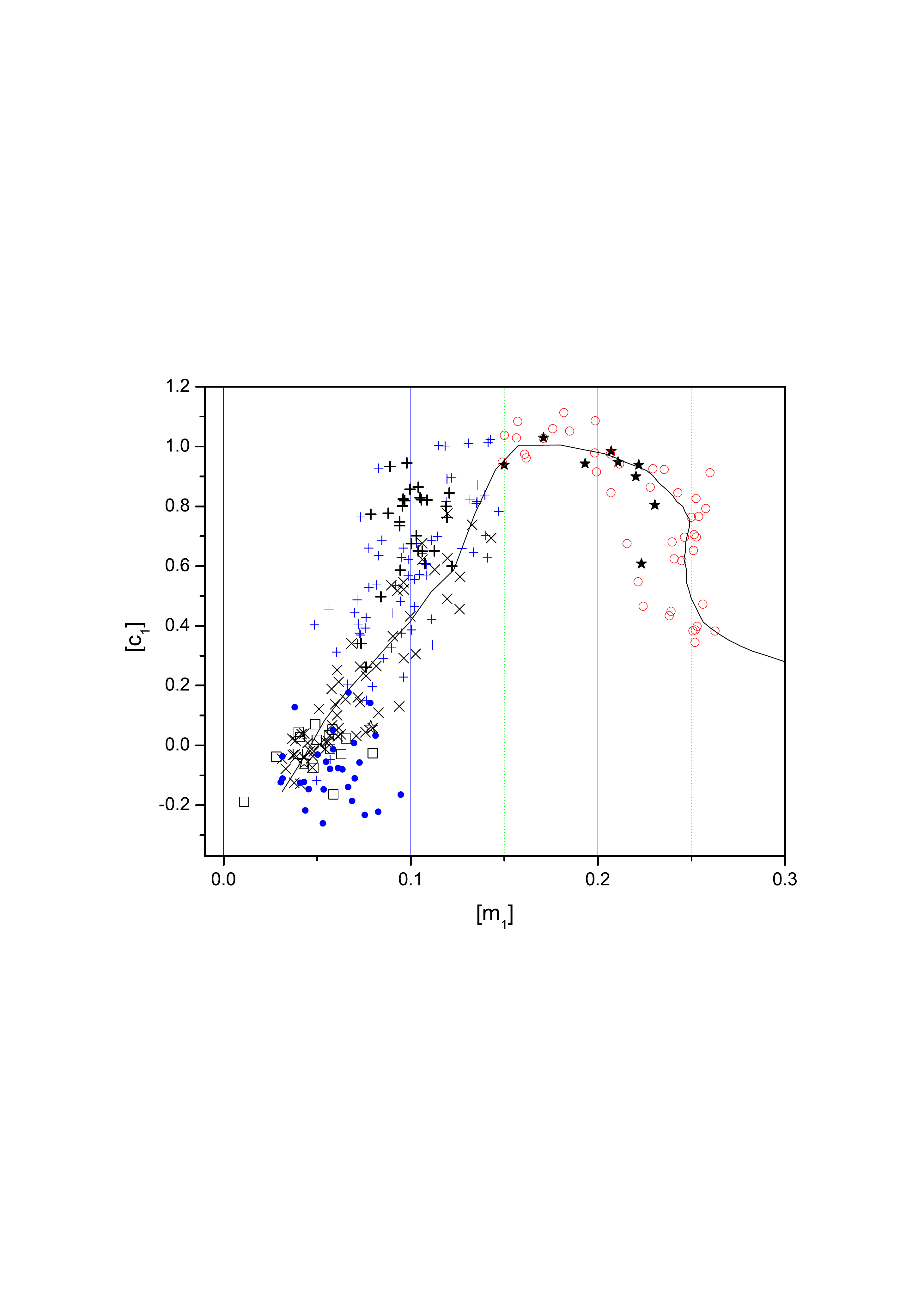}
\includegraphics[width=15pc]{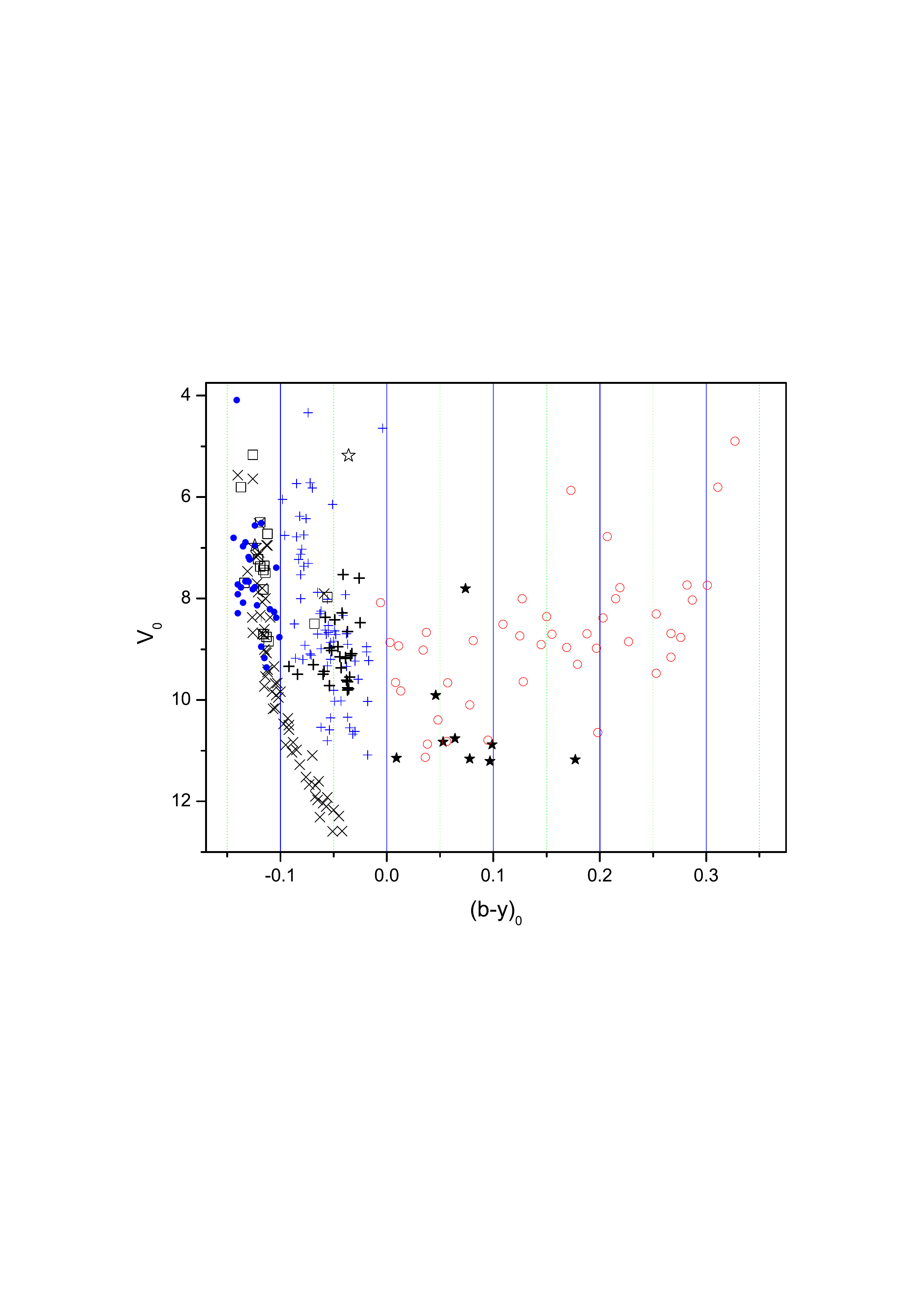}
\includegraphics[width=15pc]{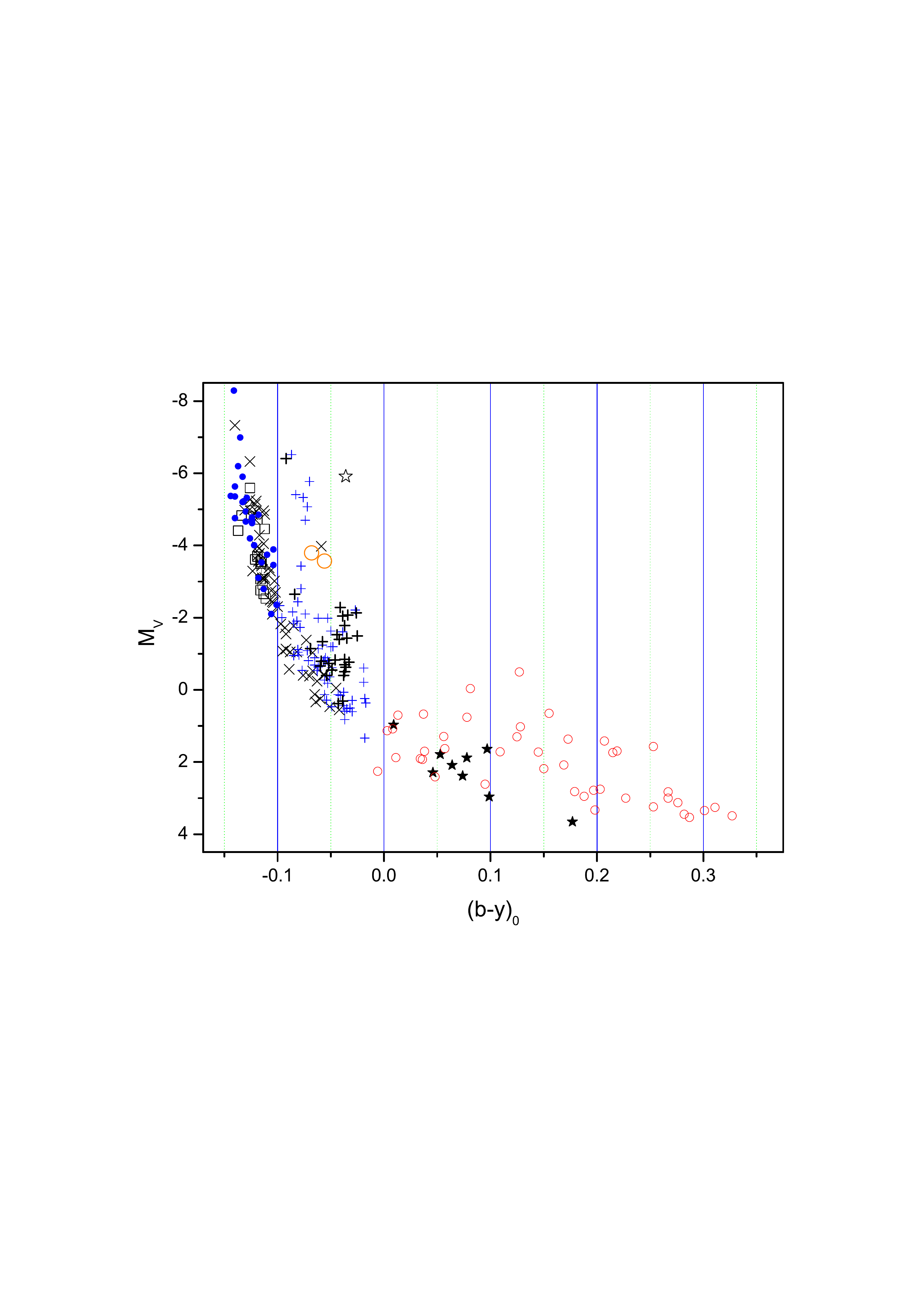}
\caption[] {All stars with calculated distances and color excesses (see Table
  1). The sample is separated into subgroups as follows: \Lod\ 112 group (11
  stars) and  \Lod\ 112 cluster (5 stars): open squares; NGC
  3293 (62 stars): x symbols;  NGC 3114 (29 stars): thick + symbols; IC 2581 (2 stars): open-star symbols; Lod$\acute{\mathrm e}$n 46
  (9 stars): filled star symbols;  blue MS field stars (29): filled circles;
  blue evolved field stars (66):  + symbols; field stars of  A-F-G types (43): open circles. \label{fig2} See the electronic edition of the PASP for a color version
  of this figure.}
\end{center}
\end{figure*}

To infer the physical stellar parameters from the photometry, the spectral and
luminosity classifications (MK) were extracted from the SIMBAD database and
used in conjunction with the classification \BrC1 vs. \BrM1 diagram
\citep{str66} to study the spectral content of the sample. The sample
contained about 40 A-F-G type stars of luminosity classes III-II-Ib, which
were clearly separated on the \BrC1 vs. \BrM1 diagram. Since for these
types a reliable estimate of distance is not available at present in the
\uvbyb\ system, they were excluded from the following analysis. The location
of the rest of the stars on the classification \BrC1 vs. \BrM1 diagram  shows
a fair agreement with the SIMBAD MK (Fig.~\ref{fig2}, top
right panel).

The procedure applied here to obtain the color excesses and stellar
distances for the O and B stars in the sample is described in detail in
\citet{kal00}. The color excesses for LC III, IV and V are obtained via Crawford's (1978) calibration. The calibration by
\citet{kilw85} is used for LC II, Ib, Iab and Ia.  We used R=3.2
and \EBV=\Eby/~0.74 to obtain $V_0$. The calibration by \citet{bal84} is
utilized for all O-B9 stars to derive the \Mv values. Since we are
dealing with early spectral types, the presence of emission lines in the
stellar spectra is the largest source of error in the calculated absolute
magnitudes.  However, the $\beta$ vs. \C0 diagram (not shown here) reveals
that very few O and B stars in the sample deviate from the main sequence and have
photometry affected by emission.  For all stars with observed $\beta$ values
outside the limits of the \Mv calibration, and all known emission-line
stars, $\beta$ calculated from \C0 was used to obtain \Mv 
\citep[cf. for details][]{bal94,kal00}. Note that this procedure yields
distance in excellent agreement with the recalculated $Hipparcos$ data \citep{kal07}. Recently \citet{kal11}  presented a comparison between the
$Hipparcos$ and \uvbyb photometric \Mv quantified as a function of the spectral
sub-type in the B0-B9 range. This comparison was done for field stars with
good-quality $Hipparcos$ parallaxes that are subject to relative errors of
less than 10\% and shows that the  agreement is in place over
the entire B-type spectral range.

For the rest of the sample we follow \citet{cra75a,cra79} and separate the
A2-F2 stars from the F2-G2 stars according to their $\beta$ indices. All
stars with $\beta$ in the range 2.55 to 2.7 we refer to as F-type and all
stars with $\beta$ between 2.7 and 2.9 we consider to be A-type.  This
photometric classification agrees very well with the SIMBAD MK types. To
derive the individual color excesses and absolute magnitude for these stellar
types, we apply the calibrations of \citet{cra75a,cra79}. The calibration of
\citet {hil83} is applied to the A0-A2 stars in the sample, which
photometric classification based on the \BrC1 vs. \BrM1 diagram is also in
excellent agreement with the SIMBAD MK types.  The photometric data and
the derived stellar parameters are summarized in Table~1 which includes the
stellar identifications, followed by the galactic coordinates, MK type and
\uvbyb\ photometric data, color excess and dereddened photometry, calculated
absolute magnitude and true distance moduli. The expected uncertainties in
$M_V$ are of the order of $\pm\,0.3$ mag for O and B types of LC III-V and
for A-F V-IV types, and $\pm\,0.5$ mag for B-type super-giants.  An
uncertainty of $\pm\,0.3$ mag in $M_V$ propagates to an asymmetric error of
$-13$\% to +15\% , and uncertainties of $\pm\,0.5$ mag result in $-21$\% to
+26\% error in the derived distances. Since the photometry used in this
  paper comes from different sources, the homogeneity of the sample is an
  important issue. Comparisons of existing \uvbyb data-sets collected by
  various authors for the field of Carina Spiral Feature present in general a
  good agreement (see \citet{kaletal00}). The latter authors estimated that the
  uncertainty in the calculated stellar distances due to possible systematic
  deviations in the existing photometric data should not exceed 3--5\%.

Note, that in the \uvbyb system both the color excess and absolute
  magnitude calculations do not rely on a precise determination of spectral
  type, since the calculations are carried on in the same way for types from O
  to B9, A0-A2, A2-F2, and F2-G2. Inspecting the reddening free \BrC1
  vs. \BrM1 diagram is sufficient to identify and resolve spectral type
  misclassifications, but we did not notice any for our sample. However a
  great care was taken to resolve all cases of suspected luminosity class
  misclassification since different calibrations are used to calculate the
  color excesses for different LC types, which is especially important in the
  O-B9 spectral range. To ensure as proper an LC classification as possible,
  the database was divided into groups according to the LC available in SIMBAD
  and each group was considered separately. The reddening free \BrC1 vs. \BrM1
  and $[c_1]$/$\beta$ diagrams (not shown here) built for each LC were used to
  examine for possible LC misclassifications and also for stars with H$_\beta$
  emission. The individual sources of spectral classification were also
  considered for all cases of observed inconsistencies, especially the
  catalogue of \citet{ree03} for OB stars and its updates. Again, we did not
  notice inconsistencies between the LC types listed in SIMBAD and the
  photometric classification diagrams we built and inspected, or other
  literature sources of luminosity classification involved in the
  comparisons.
 
\section{Photometry-derived results}
The diagrams color excess \Eby vs. distance moduli, $V_0$ vs. \by0 and \Mv
vs. \by0 are presented in Fig.~\ref{fig2} and are used to reveal spatially
coherent structures in the studied longitude range. Figures 3, 4 and 5 show
the stars with available distances plotted in galactic coordinates. The
backgrounds in these figures present the distributions of $H\alpha$, $^{12}$CO
(J=1$\rightarrow$0, 115 GHz), and dust infrared emission, and are described in
the captions of the figures.

There is a number of papers devoted to star formation in dense ISM in the
  surroundings of H{\sc ii} regions, both from theoretical and observational
  point of view (see, for example, \citet{cic09} and the references
  therein). According to the current theories massive stars tend to stimulate
  star formation at larger distances, but affect destructively their immediate
  neighborhood, since they tend to disrupt the parental molecular cloud. It
  could be seen in Figs.~4 and 5 that the location of the stars in the sample
  does not correlate with the dense molecular clouds represented by the
  distribution of the $^{12}$CO flux, and with zones of high reddening. On the
  other hand, the stellar distribution correlates with the location of ionized
  hydrogen in Fig.~3.  Studying the distributions of OB stars together with
  the distributions of $H\alpha$ (a tracer of ionized hydrogen), $^{12}$CO (a
  tracer for neutral hydrogen and molecular clouds), and dust infrared
  emission, could give more complete picture of the massive stellar population
  in the region, the different components of the ISM, and the interactions
  among them.

\subsection{\Lod\ 112 and IC 2581}

\Lod\ 112 is a poor cluster candidate that contains about 10 OB stars. Five of
them are included in our sample: HD 300811, HD 300813, HD 300814AB, CPD -56
3492, CPD -56 3496 (see Table 1).  These stars form a very compact group at a
(median) true DM = 11.06 ($\pm$0.28~sd; $\pm$0.12~se) and average color excess
\Eby = 0.5($\pm$0.07~sd; $\pm$0.03~se). For all groups under consideration in
this paper we will indicate both the standard deviation (sd) and standard
error (se) when providing distance modulus and color excess. When calculating
the distance to the groups, the uncertainties will be based on the standard
error in the distance modulus. Our distance estimate of $1629^{+84} _{-80}$ pc
for \Lod\ 112 is significantly smaller than the presently adopted 2500 pc (see
for example WEBDA database). The 2500 pc distance is based on the V
  vs. B-V diagram of the cluster \citep{kar05}. Since all known possible
  members of \Lod\ 112 are found in the upper part of the MS, a determination
  based on the location of these stars on the color-magnitude diagram is
  difficult. The \uvbyb photometry allows us to obtain the distance to each
  stars and find an average distance to the cluster. \Lod\ 112 has been last
studied by \citet{lod77} who performed UBV and \uvbyb photometry of
stars in the Carina-Crux-Centaurus-Norma region suspected of belonging to poor
open clusters or associations. However, the characteristics of these cluster
candidates have not been photometrically studied based on \uvbyb
photometry. The photometric diagrams and individual stellar distances
presented here (Fig.~\ref{fig2}) indicate that one of these candidates,
\Lod\ 112, is possibly a physical group. In our sample there are 9 other early
B main-sequence stars located at that distance (HD 84361, HD 89174 (both found
at the edge of the studied field at galactic longitude approximately
$280^\circ\!$), HD 88661, HD 90102, HD 90273, HD 90288AB, HD 90615, CPD -54
3538, CPD -55 3036). Two relatively evolved stars (HD 89714 and HD 90135) are
also found in this distance range.  For all 16 stars mentioned in this
paragraph the (median) true DM is 11.12 ($\pm$0.6~sd; $\pm$0.15~se).

The open cluster IC 2581 seems to be part of the feature of OB stars mentioned
above.  The two brightest OB stars of the cluster (HD 90706 and HD 90707) have
\uvbyb data available.  They are found at an average true DM = 11.56
($\pm$0.63~sd; $\pm$0.45~se), corresponding to a distance 2051 pc (vs. 2446 pc
provided by \citet{dia03}), and have average color excess \Eby = 0.43
($\pm$0.02~sd; $\pm$0.02~se). The cluster has been previously studied by
\citet{llo69} who found DM=12.0 (2500 pc) and \citet{tur73} (DM=11.65 (for
R=3), corresponding to 2140 pc).

The field considered in this paragraph clearly stands apart from the large
H{\sc ii} features toward Car OB1 ($\eta$ Car complex) and contains several
smaller but prominent H{\sc ii} nebulosities, among which RCW 48 and RCW 49
(Fig.~\ref{fig6}, top). This field is often called the ``preceding end of the
Carina complex'' (see \citet{llo69} for example). It is difficult to judge if
the OB stars considered in this subsection are spatially connected to these
H{\sc ii} regions, or are foreground. Note that the distant cluster Wd~2 (6.7
kpc, see  4.6) is found at the center of RCW 49.  On the other hand, IC
2581, at only about 2-2.5 kpc, seems to be located at the edge of RCW 49, the
angular distance between IC 2581 and Wd~2 being about 25 arcmin. IC 2581 is
considered of intermediate age and thus should not be involved in the bright
nebulosity. However, this may not be true for the two brightest OB stars in
the cluster, considered here. Note that it is not certain that these two stars
are actually members of the cluster.  A loose constraint between 2 and 5 kpc
of the distance to RCW 49 has been derived by \citet{tsu07} from the mean
X-ray luminosity of T-Tauri stars. \citet{asc07} proposed 2.8 kpc based on NIR
magnitudes and colors of RCW~49 sources on the Henyey track. 

For all stars considered in this subsection proper motions are available
(Fig.~\ref{fig6}, bottom), and are similar for the majority of them. Thus,
based on distances and proper motions, we suggest a new OB association at
coordinates $282^\circ\!<l<285^\circ\!$, $-2^\circ\!<b<2^\circ\!$,
connected to the compact \Lod\ 112 group and containing HD 90706 and HD
90707 (probable members of IC 2581). The (median) true DM for all 18 stars
studied in this subsection is 11.13 ($\pm$0.59~sd; $\pm$0.14~se),
corresponding to $1682^{+113}_{-104}$ pc. Having in mind the independently
obtained constraints and estimates of the distance to RCW~49, it is difficult
to judge whether this feature of OB stars is connected to the H{\sc ii}
nebulosities seen in this direction or is foreground.

\subsection{HR Car}

HR Car (HD 90177) is a luminous blue variable known to undergo slow irregular
spectrophotometric variations of about 1.5 mag over timescales of months
\citep{car79}. The most recent attempt to obtain the distance to HR Car
\citep{vangen91} is based on the reddening-distance method of field stars and
yields $5\pm1$ kpc. This estimate matches well the 5.4 kpc distance to the
Carina arm obtained kinematically \citep{hut91}.  Other existing distance
estimates are based on the assumption that HR Car belongs to the Carina
complex at the canonical distance of 2.5 kpc \citep{vio71}.  However, the line
of sight to the Carina complex is tangential to the Carina spiral arm, so the
luminous stars seen in this direction may have a very large range in distance
(see for example \citet{kal10}). The $Hipparcos$ distance obtained via the
revised parallaxes \citep{vanL07} is $592^{+577}_{-194}$ pc, thus locating the
star much closer than any other estimate. We stress however that HR Car lies in a relatively crowded region and this could affect the accuracy of the $Hipparcos$ parallax.

The visual magnitude of HR Car in general varies between 7.6 and 8.6 mag
\citep{par00}. The measured $\beta=2.392$ indicates that the star is observed
in emission and a $\beta$ index obtained via \C0 should be used when
calculating the absolute magnitude (see \citet{kal00}). This yields \Mv=-8.3
mag. The emission usually does not affect the color excess calculation. Here
we adopt \Eby=0.912 and a visual magnitude of 8.076, which provides a distance
modulus 12.37, corresponding to 3 kpc (see Table 1). This is the second most
reddened star in our sample. We estimate $A_V$=3.96 mag (for $R$=3.2).

HR Car is variable both in photometry, spectral type and luminosity
  class. The available MKK classifications indicate spectral type from B2 to
  B9 and luminosity class I.  Since the \by0 vs. \C0 relations used for
  reddening determination are identical for LC I-II near to the upper part of
  the MS \citep{kilw85}, a variable luminosity class should not influence the
  obtained color excess. As previously mentioned, both color excess and \Mv
  determination do not depend of the spectral subtype. In this sense, the
  variability in spectral classification of HR Car should not influence the
  obtained distance.

The distance and absolute magnitude of HR Car are important for a variety of
reasons, like the theoretical interpretation of this stellar type
\citep{vangen91}, studying the spatial distribution of  dust around the
star \citep{uma09}, etc. Apparently an ambiguity in both estimates is still
present. However, despite of the peculiarity of this star, our method based of
\uvbyb photometry provides reasonable estimates of both quantities and
indicates that the currently accepted distance of 5 kpc is overestimated and
should be applied with caution. On the other hand, the \Mv estimate obtain here is in
agreement with the one derived by \citet{vangen91}.

\subsection{RCW 45 and CPD $-$55 3036}

RCW 45 (BRAN 295) is a rather isolated H{\sc ii} region (Fig.~\ref{fig6}, top),
located at coordinates $l = 282.13^\circ\!$, $b = -0.11^\circ\!$, with
angular size of 16 arcmin and radial velocity \Vlsr = $-$9.8 km s$^{-1}$.
RCW 45 is included in the star-forming field Avedisova 2297, at an accepted
distance 6500 pc and angular size 30.25 arcmin.

The distance estimate of 6500 pc is based on the survey of H{\sc i} 21-cm emission
in the southern Milky Way by \citet{mcc00}. These authors have detected two
large shells in the interstellar neutral hydrogen near the Carina tangent,
centered at ($l,b$)=(277,0) and ($l,b$)=(280,0) that share a common line of
sight. The center velocities are $\sim $36 km s$^{-1}$ and $\sim $59 km
s$^{-1}$, which puts the shells at kinematic distances of 6.5$\,\pm\,0.9$
kpc and at 10 kpc, respectively.  GSH 277+00+36 can be classified as a
supershell on the basis of its large size and expansion energy. The above
authors find evidence for molecular clouds along the supershell's edges,
indicating that a star formation may have been initiated by the supershell's
expansion.  They suggest that the prior interpretation of this large void as
an interarm region is inappropriate on the basis of the supershell's chimney
and shell-like morphology. The shells should be rather considered interarm
voids, as previously suggested by Grabelsky et al. (1987).

The relation of RCW 45 to this supershell is not clear. In their catalog of
candidates for Galactic worms (the walls surrounding the superbubbles)
\citet{koo92} noted that the H{\sc ii} regions RCW 45 and RCW 46 lie at the
base of GW 281.5+1.5, which place them in the eastern edge of the shell
\citep{mcc00}. Note that RCW 45 and RCW 46 do not overlap
with the worm but are located at the plane just below the worm candidate
(table 5 of \citet{koo92}).  Also note that RCW 45 is marked with a question
mark in their Table 5. All this means that accepting the supershell distance
of 6.5 kpc as a distance to RCW 45 has to be done with caution.

CPD $-$55 3036 (LS 1448; $l = 282.166^\circ\!$, $b = -0.025^\circ\!$; \Vlsr 
not available) is located in the direction of RCW 45 and is the most
reddened star in our sample. Using UBV photometry, \citet{den77}
obtained \EBV = 1.62 (\Av = 5.18 for R = 3.2) and a distance of 1380 pc.
Based on UBV$\beta$ photometry \citet{wra80} obtained \Av = 5.1 mag and
DM = 9.2 (700 pc). The star is included in the extensive $uvby$ photometric
study of the Luminous Stars in the Southern Milky Way by Kilkenny \& Whittet
(1993). Kilkenny (1993) presents $\beta$ photometry of the star and obtains
\Av = 5.14 and a distance of 1200 pc. In this paper we determine \Av = 5.36
mag and DM = 10.21 mag, which corresponds to a distance of 1101 pc.

The high interstellar extinction toward CPD $-$55 3036 and its proximity to
RCW 45 in terms of Galactic coordinates may indicate a possible
relation. However, the uncertainty in the distance of RCW 45  discussed above 
and the lack of radial velocity measurements for the star would make at this
point further conclusions preliminary.

\subsection{\Lod\ 46}

\citet{lod79} performed \uvbyb\ of 15 stars of the cluster candidate in field
46 (\Lod\ 46) and estimated a distance of 1.24 kpc and \EBV = 0.22
mag. \Lod\ 46 is represented in our sample by 9 stars of spectral class
A0-A2. We determine a true median DM = 8.67 ($\pm\,$1.42~sd; $\pm\,$0.47~se),
corresponding to a distance 542 pc, in good agreement with the adopted
distance of 540 pc (see WEBDA database).  The relatively large spread in
individual distances for these stars may be due to the fact that the \uvbyb
system is not very suitable for the A0-A2 spectral range. 

\subsection{NGC 3293 and NGC 3114}  
These clusters are studied in details by various authors. Here we present
only  revised distances and color excesses in order to provide homogeneous
estimates for all groups with \uvbyb photometry in the  field under consideration.
 
NGC 3293 is a bright open cluster embedded in an emission nebula. \citet{sho80} obtained \uvbyb photometry of a significant amount of
cluster members and calculated true DM=12.75 (3.55 kpc). This estimate however
has been based on a preliminary \Mv-$\beta$ calibration, later found to
overestimate the brightness, and was corrected to 11.95$\pm$0.1 mag (2455 pc)
\citep{sho83}.  Our estimate is based on 61 stars with available \uvbyb
photometry and yields a true (median) DM = 12.15 ($\pm$0.43~sd; $\pm$0.05~se), 
corresponding to a distance of 2691 pc. A distance of 2373 pc is provided in the
\citet{dia03} catalog, while \citet{kar05} estimate 2471 pc. Although a fair
agreement exists between our estimate and the most recently published
distances, our result locates the cluster some 250 pc farther than currently
accepted.  The stars are uniformly reddened with average color excess \Eby =
0.20 ($\pm$0.05~sd; $\pm$0.01~se).

NGC 3114 has been observed in the \uvbyb system by \citet{sch82} and
\citet{sch88}. In this cluster, there are 29 stars earlier than B9 with \uvbyb
data presently available. The cluster is located in a crowded low-reddened
field, thus complicating the separation of cluster members from field
stars. \citet{car01} obtained UBVRI photometry of more than 2000 stars near
the center of the cluster. They found this region to be heavily contaminated
by field stars and separated two populations: several low reddened stars which
are presumably cluster members, and field stars having larger reddening.  All
stars in our sample are low reddened. We estimate an average color excess \Eby
= 0.053($\pm$0.01~sd; $\pm$0.009~se), in agreement with \citet{sch82} and
\citet{car01}. However, these stars show a significant spread in distance and
not well defined MS on the $V_0$ vs. \by0 diagram (Fig.~\ref{fig2}). After
excluding one star with very large distance, the remaining 28 stars yield a
median true DM = 10.06 ($\pm$0.71~sd; $\pm$0.13~se), corresponding to 1028 pc
(vs. 911 pc listed by \citet{dia03} and 1130 pc provided by
\citet{sch82}). Excluding the stars to the left of MS does not significantly
affect this distance estimate. Nineteen stars, however, appear to be nicely
grouped between distance moduli 9.7 and 10.5 mag, at an average true DM =
10.01 ($\pm$0.29~sd; $\pm$0.067~se). Thus, we provide a revised distance of
$1005\pm31$ pc to NGC 3114, which is in fair agreement with the estimate of
\citet{car01} of $920\pm50$ pc and \citet{sch88} of $940\pm60$ pc.

\subsection{Westerlund 2}
The massive, young stellar cluster Wd~2 ($l = 284.2^\circ\!$, $b =
-0.33^\circ\!$) is considered to be one of the five superclusters known in
the Milky Way. The cluster is thought to be spatially connected to the H{\sc
  ii} complex RCW~49, a remarkable infrared nebula as revealed by Spitzer
\citep{chu04}, and perhaps to the extended  TeV $\gamma$-ray source HESS
J1023-575 \citep{aha07}.   Recently, $\gamma$-rays in the MeV/GeV energy
domain have been reported from the same direction by the $Fermi$ 
collaboration \citep{abd09}. \citet{fuj09} present an analysis of the diffuse
X-ray emission of Wd~2 which may indicate a recent ($\sim 10^5-10^6$ yrs ago)
explosion of a massive star. Recently a new hard spectrum TeV $\gamma$-ray
source, HESS J1026582, was discovered by the H.E.S.S. collaboration
\citep{abr11}. \citet{ack11} showed that this emission is due to a
  $\gamma$-ray pulsar with a preferred distance of 2.4 kpc.  Since the most recent
  distance estimates to Wd~2 were established to be in the 5-8
  kpc range,  this $\gamma$-ray emission might be unrelated to Wd~2
  and this would cast serious doubt on a connection between the pulsar and the
  cluster (see \citet{rau11} for a thorough discussion).  Wd~2 is one of the
  clusters in the Galaxy for which associated molecular clouds have been
  identified \citep[Furukawa et al. 2009, see also][]{dam07}. \citet{fuk09}
discovered a spectacular jet and arc of molecular gas detected with the NANTEN
telescope in the $^{12}$CO $J = 1 - 0$ 115 MHz emission line survey
\citep{muz04}.

The total stellar mass in Wd~2 is of the order of 4500 $M_\odot$, including 12
O stars and 2 WR stars \citep{rau07}.   Due to the high (and apparently
local) extinction, most of the photometric and spectroscopic studies are
restricted to the brightest stars.  The only deep BVI CCD photometry is
presented by \citet{car04}.   The distance to Wd~2 has been a very
controversial issue (see \citet{dam07} and the references therein) and varies
between 2 and 8.3 kpc.

Distances to RCW 49 have been presented by several authors (see also
4.1). In the study of \citet{gra88} Wd~2 and RCW~49 are associated with
GMC~7, at an optical distance of 4 kpc.  \citet{rus03} determine a kinematic
distance of 4.7 (+0.6, $-0.2$) kpc to RCW~49.  On the basis of an analysis
of the CO emission and 21 cm absorption along the line of sight to Wd~2, 
\citet{dam07} argued that Wd~2 must be associated with GMC~8 of the study of
\citet{gra88}, in the far side of the Carina arm.  \citet{dam07} determined
a kinematic distance of $6.0\,\pm\,1.0$ kpc to the molecular structure toward
Wd~2,  while \citet{fur09} obtained $5.5\,\pm\,1.5$ kpc.  In their
original study, \citet{gra88} reported a kinematic distance of 6.6 kpc to
GMC~8.

Although \uvbyb photometry of the stars in Wd~2 is not available, we attempt
here a distance estimate for this cluster. The 12 O-type stars, studied by
\citet{rau07} are listed in Table~3.  Similarly to Rauw et al. we use the MK
type to obtain \Mv and $E(B-V)$, but utilizing the calibration by
\citet{deu76}.  We have tested this calibration based on a large sample of
O-B stars and found it to provide \Mv that is in agreement with the \uvbyb
photometry.  For example, the \uvbyb sample used in this paper contains six
O-type stars.  For them, the average \Mv derived by \uvbyb\ photometry is
$-5.2\,\pm\,0.26$ mag, and the average spectroscopic \Mv is $-5.08\,\pm\,0.13$
mag.  To do another check we recalculated the distance moduli of the stars of
the \Lod\ group with available spectral classification and found  a median
value of 11.19($\pm$1.46~sd; $\pm$0.55~se), which, despite of the
somehow larger spread, is in agreement with the \uvbyb estimate. 

The $E(B-V)$ and \Mv obtained here are listed in columns 7 and 8 of Table~3. 
Columns 9, 10 and 11 contain the distance moduli calculated for three values
of R (3.2, 4.2, 5.2).  The average distance moduli are shown at the bottom
of Table~3, together with the mean error, and correspond to distances of
6700 pc, 3090 pc and 1432 pc, respectively.  Since the \Av toward Wd~2 is
more than 5 mag, one could expect an abnormal value of R, but, as mentioned
by \citet{rau07}, the present photometric data does not allow us to evaluate
the reddening law toward Wd~2. Another way to test our distance estimate is to
obtain \by0 from \BV0 and plot the Wd~2 stars on the \Mv vs \by0 diagram. 
To do this the study of \citet[his Fig. 7]{cra75b} on the O-type stars was
used.  The average \by0 and \Mv values for the 12 stars are $-0.14$ and
$-5.5$ mag (obtained via the calibration of \citet{deu76}), respectively,
providing a good agreement with the location of the MS on Fig. 2, where the
\Mv values are obtained via the calibration of \citet{bal84}.   This points
out that both \Mv and \BV0 obtained from the \citet{deu76} calibration for the
Wd~2 stars are in agreement with the parameters provided by the \Stromgren
photometry.  Thus, utilizing R=3.2, we provide a new photometric distance to
Wd~2 of $6698^{+512} _{-475}$ pc.

The accurate and deep $BVI$ CCD photometry by \citet{car04} provides a
distance of $6.4\,\pm\,0.4$ kpc, also in excellent agreement  with the one
of 6.7 kpc derived here.  However, an investigation of the value
of R is clearly warranted in order to pinpoint the exact distance to the
cluster. 

In their study, Rauw et al. have used the new calibration for O-type
  stars by \citet{mar06}. In order to closely evaluate both calibrations, we
  used the database of \citet{mai04} and calculated color excess and absolute
  magnitude for all 273 stars with \uvbyb photometry. Then we obtained these
  parameters via the calibrations of \citet{mar06} and \citet{deu76}. The
  comparisons led to the following results. For LC V (96 stars) \citet{deu76}
  provides DM practically identical with the \uvbyb ones, while \citet{mar06}
  seem to slightly underestimate the DM by 0.25$\pm0.47$. For LC III (44
  stars) \citet{mar06} provides DM practically identical with the \uvbyb ones,
  while \citet{deu76} seem to underestimate the DM by 0.29$\pm0.73$. In the Wd
  2 sample, seven of the stars are of LC V and 5 stars are LC III. The
  calibration of \citet{deu76} provides DM=14.095$\pm0.43$ for the 7 LC V
  stars and DM=14.18$\pm0.77$ for the 5 LC III stars. For the same sample the
  calibration of \citet{mar06} provides DM=14.08$\pm0.53$ for LC V and
  DM=14.72$\pm0.79$ for LC III. It seems that in this particular case the
  \citet{deu76} calibration provides more consistency between LC V and LC
  III. Overall this is an acceptable difference in the average DM which yields
  to 6.7 kpc using \citet{deu76} and 7.4 kpc if \citet{mar06} is utilized. All
  of the above calculations are for R=3.2. In their study \citet{rau07} use
  R=3.1, which, since the reddening of Wd 2 is quite high, results in a
  distance of 8 kpc. All this points out that our result is actually
  consistent with that of \citet{rau07} and the discrepancy is mostly due to
  the utilized value of R.

Figure 7 presents the plots of DM vs. Galactic longitude and Galactic latitude
for all stars of this sample with calculated distances. The clusters and
  cluster candidates are shown, together with the GMC from the study of
  \citet{gra88}. For the clusters studied here we use the new estimates and
  utilize the distances listed in Table 2 for the rest of the clusters. The
  edge of the Carina arm based on the distribution of the bright OB stars
  \citep{gra70, kal10} is indicated with a solid line. The distances to GMC~3
  and GMC~7 are optical, all other GMC distances are kinematic (see
  \citet{gra88}).  Any distance larger than 3.5 kpc would place Wd~2 beyond
  the edge of the Carina arm, but a large distance is in agreement with the
  location of the edge of the arm delineated by the GMC. There is an excellent
  agreement between the newly obtained distance of 6.7 kpc to Wd~2 and the
  kinematic distance of GMC~8.

\section{Conclusions}

We present a \uvbyb photometric investigation of a number of clusters and
cluster candidates and field stars located in the Galactic plane toward the
tangent of the Carina arm. Based on the derived homogeneous distances and
color excesses of more than 260 stars of spectral types O to G, we provide
revised distances for the stellar groups and layers present in this
sample. The main findings are as follows:

1. The cluster candidate \Lod\ 112 seems to be a physical group at $1629^{+84}
_{-80}$ pc.  We found other OB stars with similar proper motions at that same
distance, and suggest a new OB association at coordinates $282^\circ < l <
285^\circ$, $-2^\circ < b < 2^\circ$. This feature appears connected to
\Lod\ 112 and containing at least the two brightest OB stars of IC 2581. It is
located at  $1682^{+113}_{-104}$ pc and probably related to the H{\sc ii}
nebulosities seen in this direction.

2.  The following parameters are obtained for the luminous blue variable HR
Car: \Mv=-8.3, $A_V$=3.96 mag (for $R$=3.2) and a distance of 3 kpc (adopting
a visual magnitude of 8.076). The currently accepted distance of 5 kpc to this
star seems overestimated.

3. The high interstellar extinction toward CPD $-$55 3036 (\Av = 5.36 mag) and
its proximity to RCW 45 in terms of Galactic coordinates may indicate a
possible relation to the nebula. The star is located at 1101 pc photometric
distance.

4.  The distance of $542^{+131}_{-105}$ pc to the open cluster \Lod\ 46 is in a
good agreement with the currently accepted value. However, one can notice that
the photometric distances obtained here show somehow a spread larger than
expected for a nearby cluster. This does not necessary mean a doubtful nature
for this cluster and can be due to the fact that the stars in \Lod\ 46 are of
spectral types A0-A2 and the \uvbyb system may not provide accurate stellar
parameters for this spectral range.
 
5. Based on 61 B-type stars with available \uvbyb photometry in NGC 3293 we find
a distance of 2691 pc.  Although a fair agreement exists between our estimate
and the most recently published distances, our result locates this cluster some
250 pc farther than currently accepted.

6. We provide a revised distance of $1005\pm31$ pc to the open cluster NGC
3114.  

7. Utilizing BV photometry and spectral classification of the known O-type
stars in Wd~2, we provide a new distance estimate of $6698^{+512} _{-475}$ pc,
in excellent agreement with recent distance determination to the giant
molecular structures in this direction. This estimate does not contradict
  with the 8-kpc distance provided by \citet{rau07} as the difference is due
  to a large extent to the utilized value of R.

\begin{acknowledgements}
  This work is supported by the National Science Foundation grant AST-0708950
  and an University of Wisconsin Oshkosh faculty development award.
  N.K. acknowledges support from the SNC Endowed Professorship at the
  University of Wisconsin Oshkosh.  V.G. acknowledges support by the Bulgarian
  National Science Research Fund under the grants DO 02-85/2008 and DO
  02-362/2008. This research has made use of the SIMBAD database, operated
  at CDS, Strasbourg, France.  We acknowledge the use of NASA's {\em
    SkyView} facility (http://skyview.gsfc.nasa.gov) located at NASA Goddard
  Space Flight Center  \citep[see][]{mcg98}. {\em SkyView} is a Virtual
  Observatory on the Net generating images of any part of the sky at
  wavelengths in all regimes from Radio to Gamma-Ray. We acknowledge the use
  of the Southern H-Alpha Sky Survey Atlas (SHASSA), which is supported by the
  National Science Foundation \citep{gau01}. We are thankful to an anonymous
  referee for many valuable comments that significantly improved the paper.

\end{acknowledgements}



\setcounter{table}{0}
\begin{landscape}
\begin{table}
\caption{The sample organized according to the subgroups identified in Fig.~2.}
\tiny
\vspace{0.1in}
\begin{tabular}{llrrrrrlccrrrrcrrr}
\tableline\tableline

	&		&		&		&		&		&		&		&		&		&		&		&		&		&		&		&		&		\\

ID	&         	l\d	&	b\d	&	V	&	b-y	&	\m1	&	\c1	&	$\beta$	&	\BrC1	&	\BrM1	&	MK	&	\by0	&	\C0	&	\M0	&	\Eby	&	\V0	&	\Mv	&	DM	\\
	&		&		&		&		&		&		&		&		&		&		&		&		&		&		&		&		&		\\
\tableline
    {\bf Lod 112}	&	--	&	--	&	--	&--		&	--	&	--	&	--	&	--	&	--	&	--	&	--	&	--	&	--	&	--	&	--	&	--	&	--	\\
HD 300814AB	&	284.7081	&	1.1309	&	9.299	&	0.337	&	-0.055	&	0.102	&	2.601	&	0.035	&	0.056	&	B3	&	-0.115	&	0.016	&	0.094	&	0.452	&	7.356	&	-3.54	&	10.90	\\
HD 300813	&	284.673         &	1.135	&	9.58	&	0.394	&	-0.073	&	0.067	&	2.602	&	-0.012	&	0.057	&	B0	&	-0.119	&	-0.031	&	0.096	&	0.513	&	7.372	&	-3.69	&	11.06	\\
LID -5603492	&	284.752	        &	1.137	&	10.86	&	0.386	&	-0.062	&	0.1	&	2.638	&	0.023	&	0.065	&	~	&	-0.116	&	0.005	&	0.104	&	0.502	&	8.701	&	-2.75	&	11.45	\\
LID -5603496	&	284.804	        &	1.14	&	10.69	&	0.318	&	-0.056	&	0.134	&	2.64	&	0.070	&	0.049	&	~	&	-0.111	&	0.052	&	0.086	&	0.429	&	8.844	&	-2.53	&	11.37	\\
HD 300811	&	284.6227	&	1.2106	&	9.88	&	0.496	&	-0.084	&	0.073	&	2.608	&	-0.026	&	0.080	&	B	&	-0.121	&	-0.044	&	0.12	&	0.617	&	7.227	&	-3.61	&	10.84	\\
{\bf Lod 112 group}	&	--	&	--	&	--	&	--	&	--	&	--	&	--	&	--	&	--	&	--	&	--	&	--	&	--	&	--	&	--	&	--	&	--	\\
HD 89714ABV	&	283.5801	&	-0.3534	&	9.04	&	0.193	&	-0.024	&	0.011	&	2.608	&	-0.028	&	0.040	&	B2Iab	&	-0.056	&	-0.036	&	0.058	&	0.249	&	7.97	&	-3.57	&	11.54	\\
HD 90135	&	282.434	        &	2.26	&	9.373	&	0.136	&	0.018	&	-0.002	&	2.6	&	-0.029	&	0.063	&	B1/B2Ib	&	-0.068	&	-0.041	&	0.085	&	0.204	&	8.495	&	-3.79	&	12.29	\\
HD 90102	&	284.063	        &	-0.358	&	8.706	&	0.178	&	-0.009	&	0.054	&	2.605	&	0.018	&	0.050	&		&	-0.116	&	-0.002	&	0.088	&	0.294	&	7.44	&	-3.49	&	10.93	\\
LID -5403538	&	282.651	        &	1.835	&	10.0	&	0.176	&	0	&	0.088	&	2.637	&	0.053	&	0.058	&	~	&	-0.113	&	0.033	&	0.095	&	0.289	&	8.757	&	-2.65	&	11.41	\\
HD 89174	&	280.2494	&	3.6794	&	7.953	&	0.173	&	-0.029	&	-0.003	&	2.581	&	-0.038	&	0.028	&	B1Ib/II	&	-0.112	&	-0.057	&	0.065	&	0.285	&	6.727	&	-4.46	&	11.18	\\
HD 90288AB	&	284.1049	&	-0.0865	&	8.148	&	-0.04	&	0.054	&	0.02	&	2.622	&	0.028	&	0.041	&B2III/IV 	&	-0.116	&	0.006	&	0.079	&	0.076	&	7.823	&	-3.06	&	10.88	\\
HD 88661	&	283.08	        &	-1.481	&	5.72	&	0.003	&	0.047	&	-0.075	&	2.448	&	-0.076	&	0.048	&	B2IVnpe	&	-0.126	&	-0.099	&	0.089	&	0.129	&	5.166	&	-5.59	&	10.75	\\
LID -5503036	&	282.166	        &	-0.025	&	11.151	&	1.106	&	-0.354	&	0.033	&	2.618	&	-0.188	&	0.011	&	B+...	&	-0.137	&	-0.203	&	0.056	&	1.243	&	5.806	&	-4.40	&	10.21	\\
HD 84361	&	280.073	        &	-3.872	&	8.351	&	0.085	&	0.012	&	0.062	&	2.507	&	0.045	&	0.040	&	B2/B3V	&	-0.114	&	0.024	&	0.078	&	0.199	&	7.496	&	-2.92	&	10.42	\\
HD 90615	&	284.301	        &	0.211	&	8.207	&	0.278	&	-0.049	&	-0.005	&	2.577	&	-0.061	&	0.043	&	B1II	&	-0.119	&	-0.08	&	0.082	&	0.397	&	6.499	&	-4.72	&	11.21	\\
HD 90273	&	284.175	        &	-0.253	&	9.075	&	0.187	&	-0.003	&	-0.126	&	2.601	&	-0.163	&	0.059	&		&	-0.134	&	-0.187	&	0.103	&	0.321	&	7.693	&	-4.83	&	12.52	\\
{\bf NGC 3293}	&	--	&	--	&	--	&	--	&	--	&	--	&	--	&	--	&	--	&		&	--	&	--	&	--	&	--	&	--	&	--	&	--	\\
LID 232930124	&	285.915	        &	0.085	&	12.61	&	0.17	&	0.04	&	0.58	&	2.8	&	0.546	&	0.096	&	B6.5III	&	-0.064	&	0.535	&	0.117	&	0.234	&	11.603	&	0.34	&	11.26	\\
LID 232930090	&	285.882	        &	0.058	&	12.04	&	0.15	&	0.07	&	0.52	&	2.74	&	0.49	&	0.119	&	B5V	&	-0.07	&	0.478	&	0.143	&	0.22	&	11.095	&	-0.37	&	11.47	\\

...	&	...	&	...	&	...	&	...	&
...	&	...	&	...	&	...	&	...	&
...	&	...	&	...	& ...	&	...	&	...	&	...	&	....	\\

\tableline\tableline

\tablecomments{Stellar identifications (HD, LID or other), followed by
  galactic coordinates, $uvby\beta$ photometric data, spectral classification (as they appear in the SIMBAD data base),
  dereddened photometry and color excess, calculated absolute magnitude and
  distance modulus. Table 1 is published in its entirety in the electronic
  edition of the PASP. A portion is shown here regarding its form and content.}
\end{tabular}
\end{table}
\end{landscape}

\setcounter{table}{1}
\begin{table}
\caption{Known clusters \citep{dia03} and \Lod\ cluster candidates in the
  field studied.}
\tiny
\vspace{0.1in}
\begin{center}
\begin{tabular}{lllrrr}
\tableline\tableline
Name                    &       $l$     &       $b$     &      Distance &       Age     &       DM      \\
                        &       deg     &       deg     &      pc       &       Myr     &       mag     \\
\tableline                        
H 90                    &       283.099 &       -1.476  &       2572    &       7.943   &       12.05     \\
NGC 3114                &       283.332 &       -3.84   &       911     &       8.093   &       9.79      \\
NGC 3330                &       284.188 &       ~3.848   &       894     &       8.229   &       9.75      \\
Westerlund 2            &       284.276 &       -0.336  &       6400    &       6.3     &       14.03     \\
Collinder 220           &       284.567 &       -0.342  &       1547    &       8.083   &       10.94     \\
IC 2581                 &       284.588 &       ~0.035   &       2446    &       7.142   &       11.94     \\
Saurer 3                &       285.096 &       ~2.999   &       9550    &       9.3     &       14.90     \\
NGC 3293                &       285.856 &       ~0.074   &       2327    &       7.014   &       11.83     \\
NGC 3255                &       286.088 &       -2.635  &       1445    &       8.3     &       10.79     \\
NGC 3324                &       286.228 &       -0.188  &       2317    &       6.754   &       11.82     \\
Carraro 1               &       286.236 &       -0.29   &       1900    &       9.48    &       11.39       \\
Collinder 223           &       286.358 &       -1.705  &       2820    &       8       &       12.25     \\
                        &               &               &               &               &         \\
\Lod\ 1                 &       281.02  &       -0.17   &       360     &       9.29    &       7.78     \\
\Lod\ 27                &       282.171 &       -0.335  &               &               &        \\
\Lod\ 28                &       282.21  &       -2.21   &       3950    &       7.3     &       12.98     \\
\Lod\ 46                &       282.6   &       ~2.07    &       540     &       9.03    &       8.66     \\
\Lod\ 59                &       283.08  &       ~3.28    &       650     &       8.45    &       9.06     \\
\Lod\ 89                &       284.35  &       ~0.87    &       380     &       8.47    &       7.89     \\
\Lod\ 112               &       284.68  &       ~1.17    &       2500    &       6.96    &       11.98     \\
\Lod\ 143               &       285.3   &       -0.89   &       600     &       8.45    &       8.89     \\
\Lod\ 153               &       285.67  &       ~0.09    &       2670    &       6.74    &       12.13     \\
\Lod\ 165               &       285.935 &       -0.365  &       1900    &       9.48    &       11.39       \\
\Lod\ 172               &       286.285 &       -2.502  &               &               &         \\
\Lod\ 189               &       286.72  &       ~2.59    &       720     &       8.64    &       9.28     \\

\tableline\tableline
\tablecomments{The object's name and galactic coordinates are given in the
  first three columns, followed by distance and age (where available), and true distance moduli. All data in this table are adopted 
from the catalog of \citet{dia03}.}
\end{tabular}
\end{center}
\end{table}

\begin{table*}
\caption{Distance and color excess for the O-type stars in Westerlund 2.} 
\tiny
\vspace{0.1in}
\begin{tabular}{llllllllrrr}
\tableline\tableline

MSP No  &       MK      &       $B$       &       $V$       &       $B-V$
&       $(B-V)_0$       &       $M_V$   &       $E(B-V)$  &       DM (R = 5.2)
&       DM (R = 4.2) &       DM (R = 3.2) \\
\tableline
18      &       O5V     &       14      &       12.8    &       1.2     &       -0.34   &       -5.6    &       1.54    &       10.39  &       11.93  &       13.47  \\
151     &       O6III   &       15.6    &       14.33   &       1.27    &       -0.332  &       -5.5    &       1.60   &       11.50 &       13.10 &       14.70\\
157     &       O6.5V   &       15.5    &       14.14   &       1.36    &       -0.332  &       -5.3    &       1.69   &       10.64 &       12.33 &       14.03 \\
167     &       O6III   &       15.48   &       14.19   &       1.29    &       -0.332  &       -5.5    &       1.62   &       11.26 &       12.88 &       14.50 \\
171     &       O5V     &       15.83   &       14.44   &       1.39    &       -0.34   &       -5.6    &       1.73    &       11.04  &       12.77  &       14.50  \\
175     &       O6V     &       15.15   &       13.93   &       1.22    &       -0.335  &       -5.4    &       1.55   &       11.24  &       12.80  &       14.35  \\
182     &       O4III   &       15.69   &       14.43   &       1.26    &       -0.34   &       -5.6    &       1.60     &       11.71   &       13.31   &       14.91   \\
183     &       O3V     &       15.03   &       13.61   &       1.42    &       -0.34   &       -5.6    &       1.76    &       10.06  &       11.82  &       13.59  \\
188     &       O4III   &       14.6    &       13.32   &       1.28    &       -0.34   &       -5.6    &       1.62    &       10.50  &       12.12  &       13.74  \\
199     &       O3V     &       15.74   &       14.39   &       1.35    &       -0.34   &       -5.6    &       1.69    &       11.20  &       12.89  &       14.58  \\
203     &       O6III   &       14.66   &       13.22   &       1.44    &       -0.332  &       -5.5    &       1.77   &       9.51  &       11.28 &       13.05 \\
263     &       O6V     &       16.5    &       14.91   &       1.59    &       -0.335  &       -5.4    &       1.92   &       10.30    &       12.22  &       14.15   \\
        &               &               &               &               &
&               &               &       {\bf 10.78}   &       {\bf 12.45}   &
{\bf 14.13}   \\
        &               &               &               &               &               &               &               &       $\pm$0.19s.e.  &       $\pm$0.17s.e.  &       $\pm$0.16s.e.  \\

\tableline\tableline
\tablecomments{MSP identification \citep{mof91}, MK classification and $B$ and $V$
  magnitudes from SIMBAD.  $(B-V)_0$  and $M_V$ are based on the calibration
  of \citet{deu76}. The last three columns  present the distance
  moduli calculated for R = 5.2, 4.2 and 3.2. The average distance moduli and
  the mean errors are shown at the end of the table for the three values of R.}
\end{tabular}
\end{table*}

\clearpage

\clearpage

\clearpage
\begin{figure*} 
\begin{center}
\includegraphics[width=29pc]{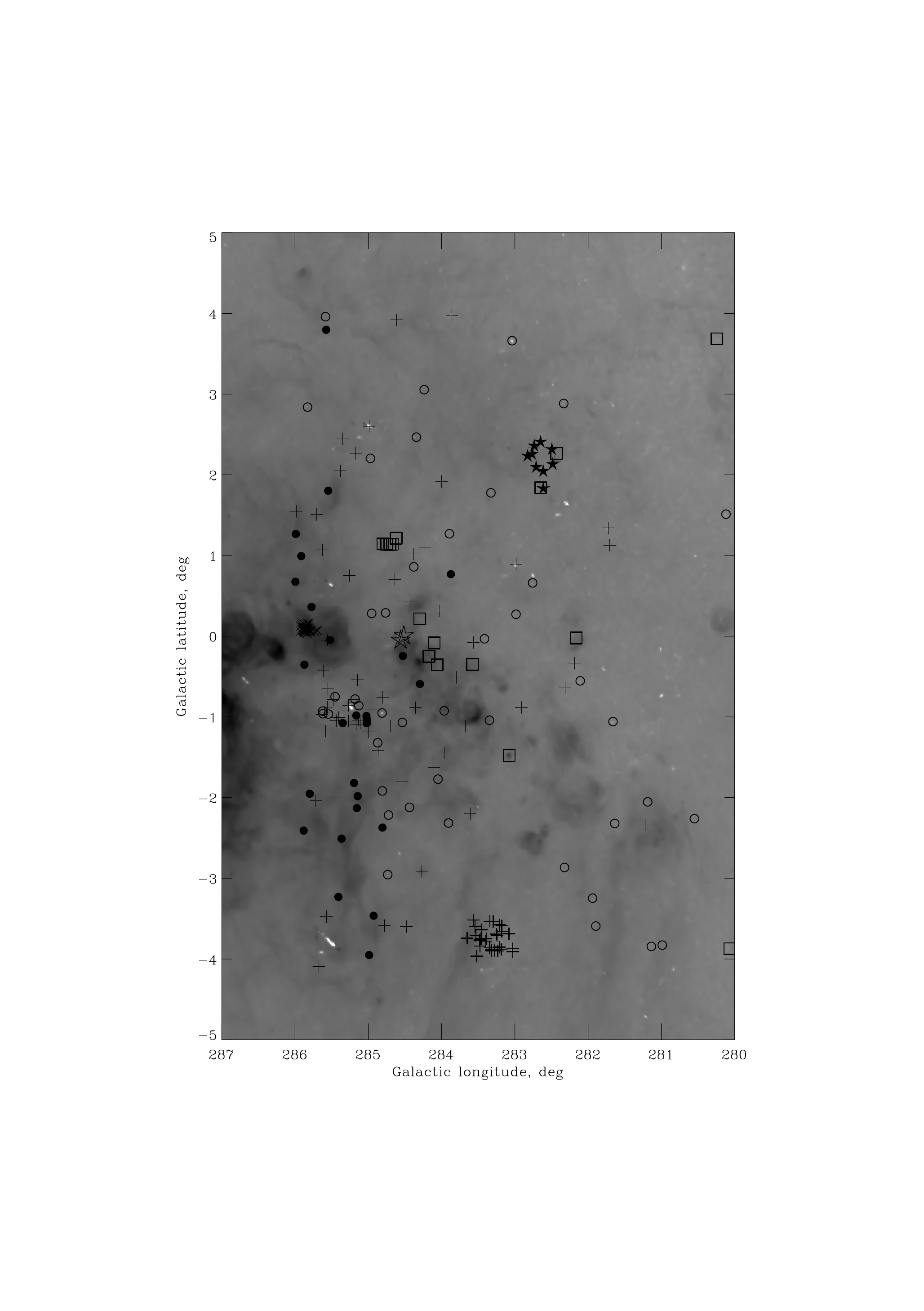}
\caption{All stars with calculated distances and color excesses overplotted
on H$\alpha$ map (smoothed to 4 arcmin resolution to remove star residuals,
\citet{gau01}), obtained via the {\em SkyView} VO interface
\citep[see][]{mcg98}. The white spots and patches in the H$\alpha$ image
are artefacts after star's subtraction. The symbols are the same as in
Fig.~\ref{fig2}.\label{fig3}}
\end{center}
\end{figure*}

\clearpage
\begin{figure*} 
\begin{center}
\includegraphics[width=29pc]{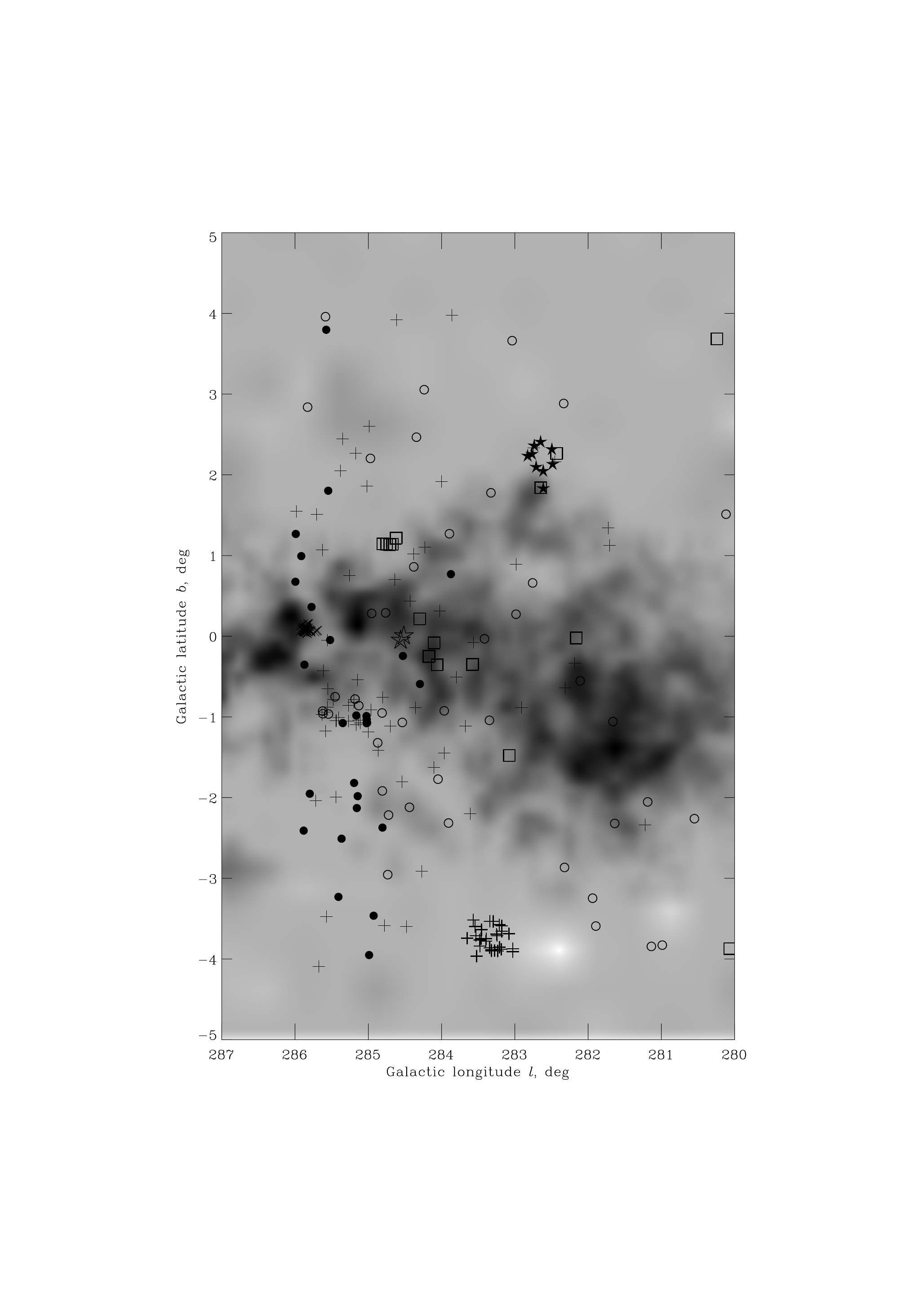}
\caption{All stars with calculated distances and color excesses overplotted on
  $^{12}$CO (J=1$\rightarrow$0, 115 GHz) map of the region \citep{dam01},
  obtained via the {\em SkyView} VO interface \citep[see][]{mcg98}. The symbols are the
  same as in Fig.~\ref{fig2}.\label{fig4}}
\end{center}
\end{figure*}

\clearpage
\begin{figure*} 
\begin{center}
\includegraphics[width=29pc]{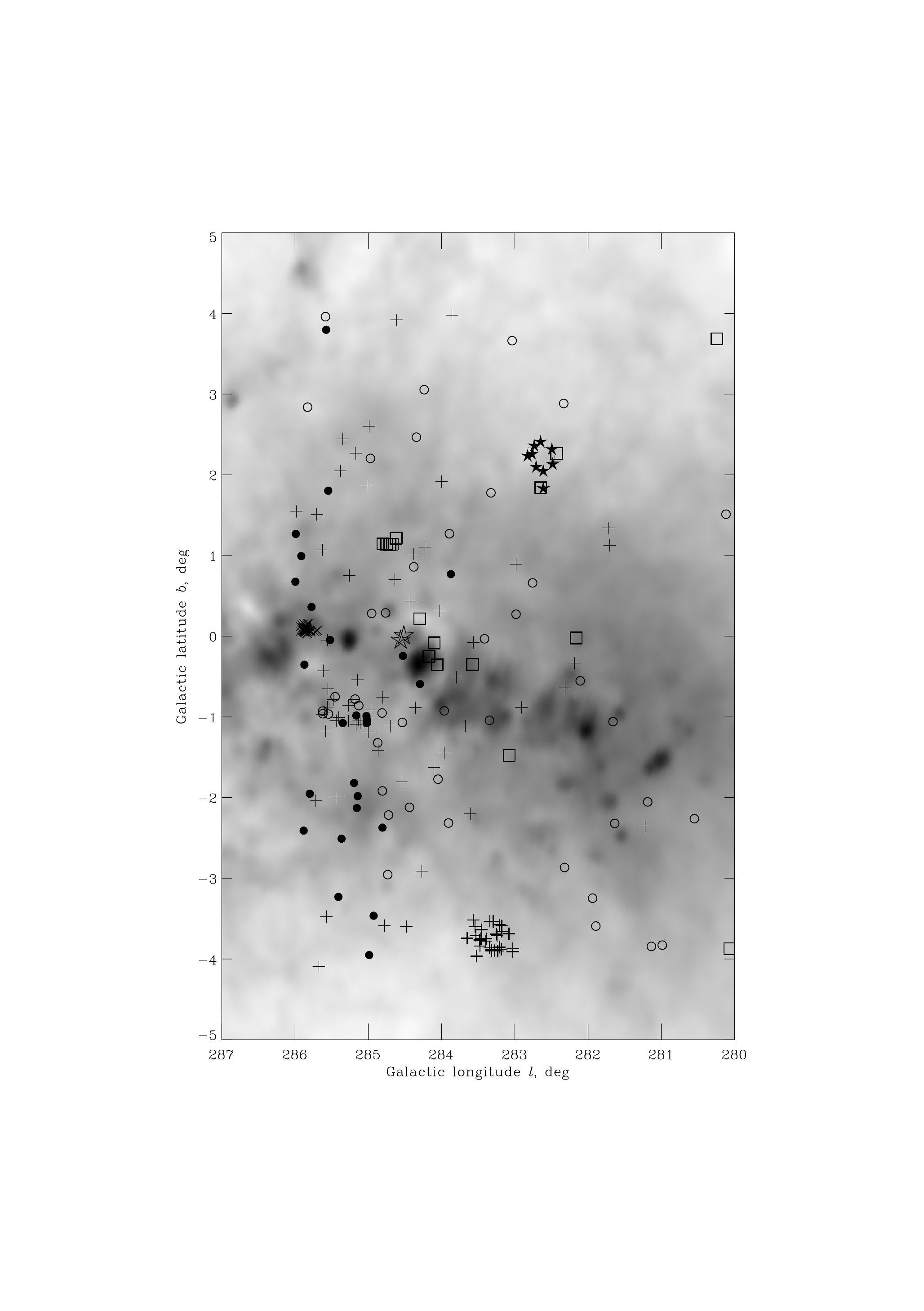}
\caption[] {All stars with calculated distances and color excesses overplotted on \EBV\ reddening map of the region \citep{sch98} obtained via the {\em SkyView} VO interface \citep[see][]{mcg98}. The symbols are the same as in Fig.~\ref{fig2}.\label{fig5}}
\end{center}
\end{figure*}

\clearpage
\begin{figure*} 
\begin{center}
\includegraphics[width=15pc]{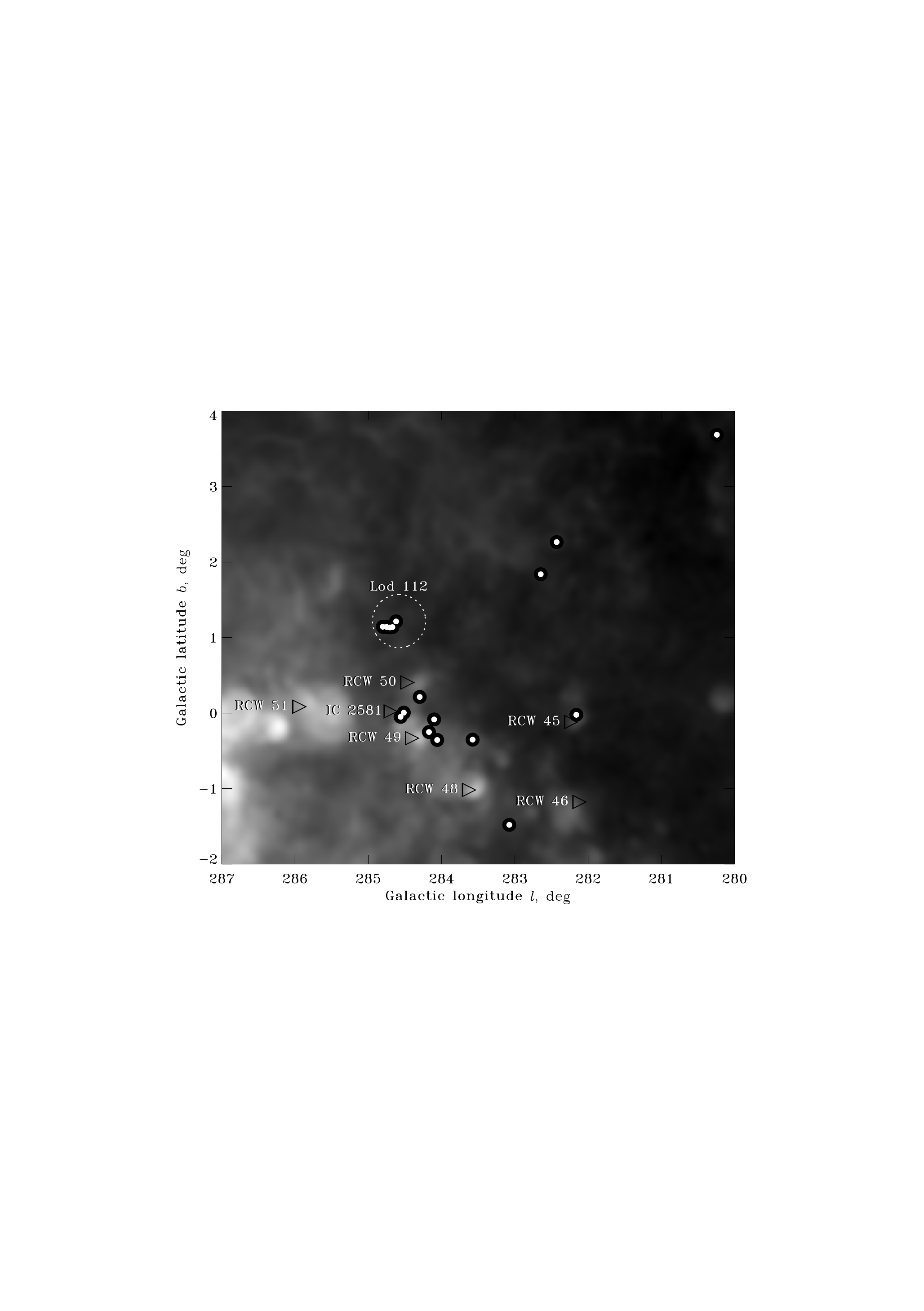}
\includegraphics[width=15pc]{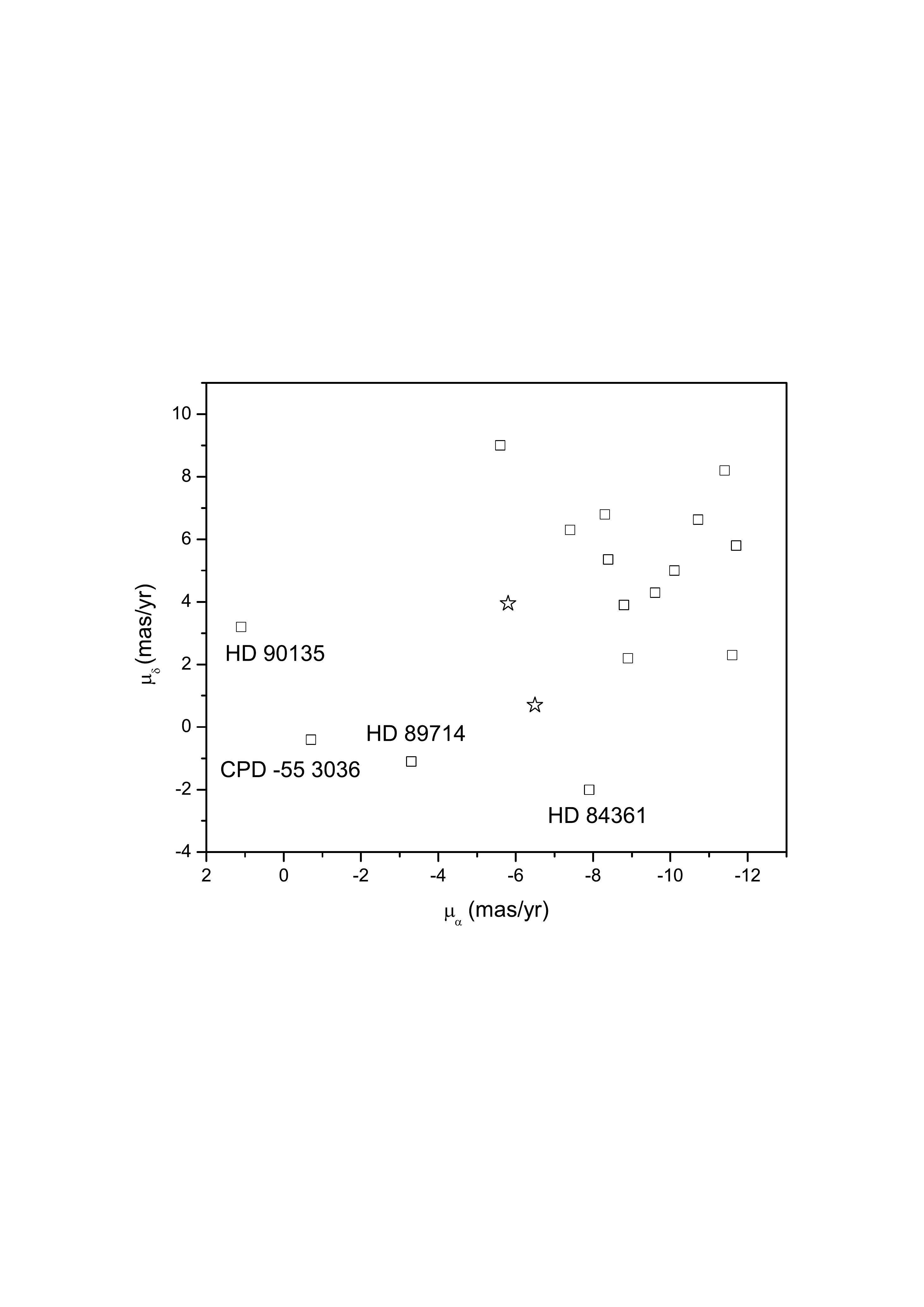}
\caption[] {(Top). The stars of \Lod\ 112 and the suggested feature of OB stars
  overplotted on the distribution of the H{\sc ii} emission \citep{fin03} obtained via the {\em SkyView} VO interface (\cite{mcg98}). The
  most prominent H{\sc ii} regions are labeled. (Bottom). Proper motions from
  the recalculated $Hipparcos$ catalog for these stars. The two possible
  members of IC 2581 are marked with asterisks. \label{fig6}}
\end{center}
\end{figure*}

\begin{figure*} 
\begin{center}
\includegraphics[width=15pc]{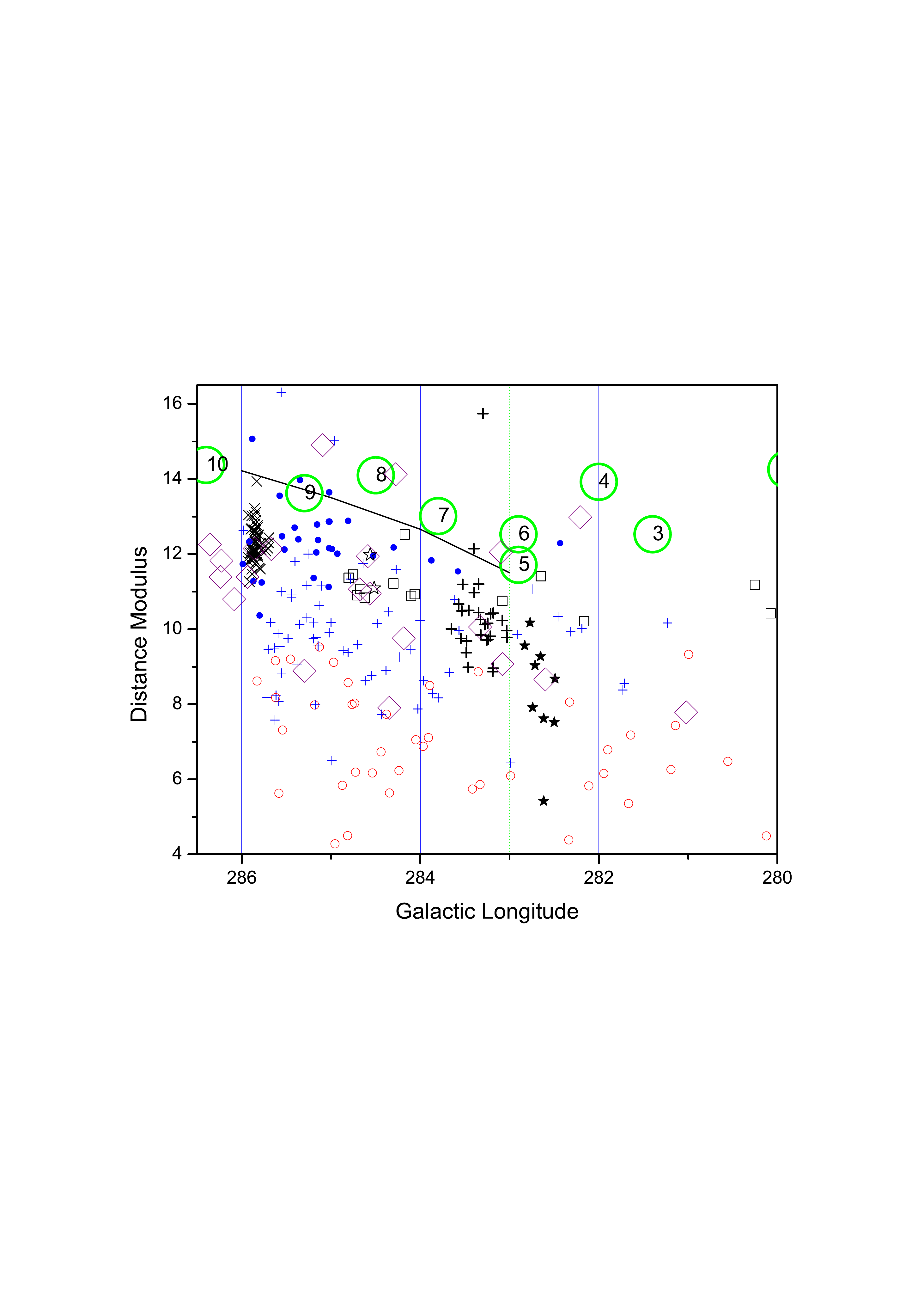}
\includegraphics[width=15pc]{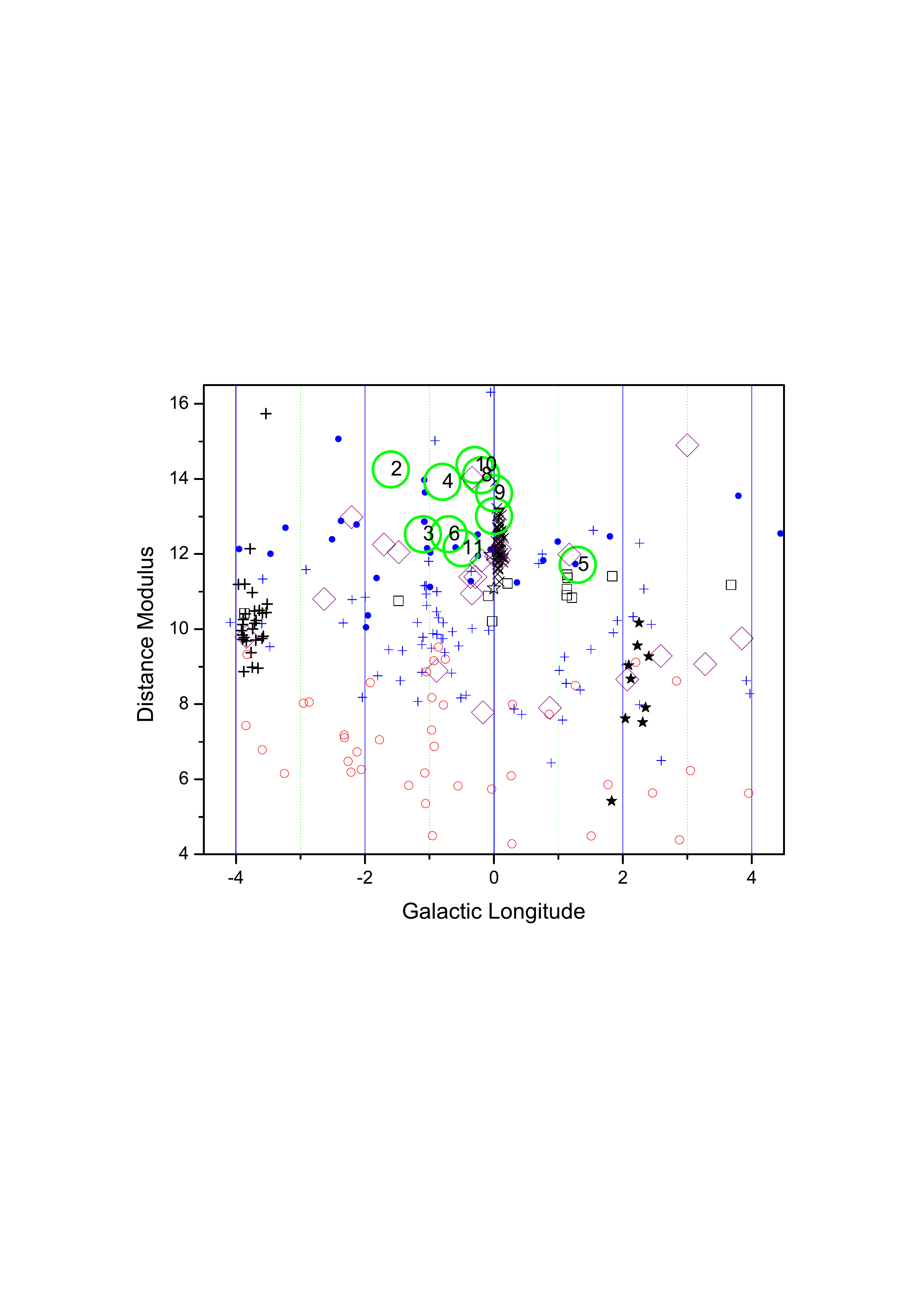}
\caption[]{(Top). Distance modulus vs. Galactic longitude for the sample
  stars (the symbols for the stars are the same as Fig. 2). The solid line is the relation obtained by Graham (1970) for the edge of the Carina
  arm.  The large open symbols (labeled) are the GMC from the study of
  Grabelsky et al. (1988). The clusters and cluster
  candidates in the field are shown with diamonds. For the clusters studied
  here the newly obtained distances are used. For the rest of the clusters the
  accepted distances (Table~2) are utilized. (Bottom). Distance moduli vs. Galactic latitude for the
  same objects.\label{fig7} See the electronic edition of the PASP for a color version
  of this figure.}
\end{center}
\end{figure*}

\end{document}